\documentclass[showpacs, aps, prb, unsortedaddress,showkeys,longbibliography]{revtex4-1}

\usepackage{amssymb}
\usepackage{latexsym}
\usepackage{amsmath}
\usepackage{graphicx}
\usepackage{braket}
\usepackage{mathdots}
\usepackage[dvipsnames]{xcolor}
\usepackage{float}

\usepackage[utf8]{inputenc}
\usepackage{gensymb}
\usepackage{hyphenat}
\usepackage{subcaption}
\usepackage{multirow}
\usepackage{graphicx}
\usepackage{bm}

\usepackage{tikz}
\usetikzlibrary{patterns,hobby}

\usepackage{amsmath,amsfonts}

\newcommand{\w}[1]{{\omega}}

\usepackage{mathtools}

\usepackage[colorlinks=true]{hyperref}

\usepackage{fancyvrb}
\usepackage{tgcursor}

\usepackage{xcolor}

\newcommand{\fref}[1]{Fig.~\ref{#1}}

\begin{document}

\title{The Atomistic Green's Function method for acoustic and elastic wave-scattering problems}

\date{\today}
\author{Hossein Khodavirdi$^a$, Zhun-Yong Ong$^b$, Ankit Srivastava$^a$}
\thanks{Corresponding author}
\affiliation{a. Department of Mechanical, Materials, and Aerospace Engineering Illinois Institute of Technology, Chicago, IL, 60616 USA,}
\affiliation{b. Institute of High-Performance Computing (IHPC), Agency for Science, Technology and Research (A$*$STAR), 1 Fusionopolis Way, $\#$16-16 Connexis, Singapore 138632, Republic of Singapore.}

\begin{abstract}
In this paper, we present a powerful method (Atomistic Green's Function, AGF) for calculating the effective Hamiltonian of acoustic and elastic wave-scatterers. The ability to calculate the effective Hamiltonian allows for the study of scattering problems in infinite systems without the introduction of any artificial truncating boundaries such as perfectly matched layers or Dirichlet to Neumann (DtN) maps. Furthermore, the AGF formalism also allows for the efficient calculation of the Green's function of the scatterer as well as all relevant scattering metrics including reflection and transmission ratios. The formalism presented here is especially suited to scattering problems involving waveguides, phononic crystals, metamaterials, and metasurfaces. We show the application of the method to three scattering problems: scattering from a slab (1D), scattering from a finite phononic crystal (1D), and scattering from defects in a waveguide (2D).

\end{abstract}

\maketitle

\section{Introduction}\label{sec:intro}

In this paper, we consider the broad problem of wave scattering from a finite scatterer into an infinite environment from the perspective of the Atomic Green's Function (AGF) method. The AGF method is appealing as it results in the Green's function of the scatterer, reduces the infinite problem to a finite problem without any arbitrary spatial truncations, and provides direct expressions for the efficient calculation of all relevant scattering parameters. The scattering problem has a long history of research, with the basic underlying ideas summarized in classic resources~\cite{achenbach1984wave,de2001handbook}. The problem finds application in many areas, too numerous to summarize here but includes guided-wave scattering~\cite{srivastava2010quantitative,song2005ultrasonic,zhuang1997axisymmetric,huthwaite2016guided}, seismic scattering\cite{wu1989scattering,sato2012seismic}, medical tomography\cite{arridge1999optical,dean2019acoustic}, calculation of radar\cite{youssef1989radar,penttila2006radar} and sonar\cite{gaunaurd1985sonar,peterson1976acoustic} cross-sections, etc.

A major concern in such problems is the process by which the infinite domain (the environment) is accounted for in the numerical scheme. A whole class of techniques, termed wave-based methods~\cite{desmet1998wave,deckers2014wave}, aims to tackle this by expressing the solution in the environment using known basis functions which automatically satisfy the associated wave equation there. In these methods, there is no artificial truncation of the environment domain, however, they are known to suffer from slow convergence and ill-conditioned matrices\cite{deckers2014wave,antunes2018numerical}. A separate class of methods depends upon truncating the infinite environment domain using an artificial boundary, thus making the problem amenable to computations. In such methods, it is of paramount importance to determine the appropriate boundary conditions which would prevent spurious reflections from the artificial boundary from polluting the solution in the interior. An exact boundary condition which accomplishes this -- called the Dirichlet-to-Neumann (DtN) map -- was discovered by Keller and Givoli in their landmark paper \cite{givoli1990non}. Even though the DtN map is exact, it is computationally challenging due to its non-local nature. To mitigate this, a whole host of approximate but local boundary conditions have been proposed over the years. Among them, Engquis and Majda \cite{engquist1977absorbing} modified the Sommerfeld radiation condition and expressed the outward normal derivative of the scattered field as approximate local differential operators. Also in other studies \cite{bayliss1980radiation,feng1983finite}, asymptotic expansions of the scattered field in the far field, or combination of a Green's function and integral approximation on the boundary were used to make other local approximate boundary conditions. Yet another approximate alternative route, more popular in FEM programs, is the application of perfectly matched layers (PML)\cite{rylander2004perfectly} just outside the artificial truncation. The purpose of PMLs is to absorb the outgoing waves at the artificial truncation and dissipate away the energy using fictitious dissipation terms. Keeping in mind the necessity of non-reflecting boundaries, different variations of finite element-based methods were also applied in the field of structural inspection. More specifically, frequency-based FE models like Spectral Finite Element (SFE) \cite{mahapatra2003spectral}, and Wave Finite Element method (WFE) \cite{ichchou2007guided} have been used to derive wave dispersion \cite{VAZIRIASTANEH201627} and scattering in elastic waveguides with flaws \cite{ZHOU20102099}.\par 

A tangentially related set of techniques which also does not depend upon artificial truncation is the Boundary Element Method (BEM) \cite{langdon2006wavenumber,chen1992boundary} and it exploits the fundamental solutions to the wave equations to solve the radiation/scattering problem~\cite{ BOUCHON2007157,perrey2004plane}. BEM is also an effective method for non-destructive evaluation of cracks using Rayleigh waves, as it allows for the solution of integral equations derived from the Betti-Rayleigh reciprocity theorem \cite{ZHANG1988365}. BEM techniques suffer from hypersingular integrals \cite{chen1999review} compared to the Finite element method. To overcome the singularity issue, special numerical integration techniques, such as the Nyström method \cite{tausch2019nystrom} or the collocation method \cite{gomez2016variational}, can be used to evaluate the integral equation more accurately. Additionally, special regularization techniques can be applied to the integral equation to avoid the hypersingularities~\cite{granados2001regularization,gu2016numerical}. Compared with the Finite Element Method (FEM), the system matrices resulting from the BEM machinery tend to be smaller in size, but the computational cost for assembling them can be higher since they are denser~\cite{laforce2006pe281}. However, sparse matrix techniques \cite{bunch2014sparse} and iterative methods \cite{ tijhuis1989iterative} have been used to reduce the memory requirements.

Scattering problems have also gained prominence in the metamaterials literature. This began with research in the development of cloaks~\cite{leonhardt2006optical,al2005achieving,leonhardt2006notes,Norris2015,norris2008acoustic,norris2011elastic,norris2012hyperelastic,srivastava2015elastic,srivastava2021causality} which seeks to minimize the total scattering cross-section of a finite region. The field of metasurfaces deals exclusively with controlling wave-scattering through the design of patterned interfaces \cite{assouar2018acoustic,zhao2013manipulating,xie2014wavefront}. Yet another recent set of related applications has been in the area of semi-infinite metamaterials~\cite{Srivastava2017EvanescentApproach}, with the elucidation of exotic phenomenon like exceptional points in conservative systems~\cite{mokhtari2020scattering,lustig2019anomalous}. Given the vast and very disparate literature on wave scattering, a unifying understanding could be made by approaching the problem through the perspective of open systems~\cite{Livsic}.  An open system, in a general sense, consists of a finite subsystem (scatterer) with discrete eigenvalues (levels of energy) which is coupled to an environment possessing a spectrum which is continuous~\cite{boundscattering}. A major aim of the theory of open systems is the derivation of the effective Hamiltonian of the scatterer which is a finite dimensional non-hermitian matrix which encapsulates all the scattering behavior of the scatterer. In earlier works~\cite{Deymier2017}, some scattering problems in discrete mass-spring systems are solved using the interface response theory. This formalism allows the calculation of the Green's function of a perturbed system in terms of the Green's functions of unperturbed systems. Although both methods are based on the calculation of Green's functions, the method of interface response theory follows a complicated and relatively long path. On the other hand, using the open system point of view and calculating the effective Hamiltonian present a simpler and physically more understandable approach in solving wave scattering problems.

The notions of the open system and effective Hamiltonian are closely linked to the Green's function framework. In condensed matter physics, there exists a large corpus of Green's function-based techniques used to study the scattering of quasiparticles such as phonons, which are the wavelike excitations of the crystal lattice, by local defects. In particular, considerable advances in the treatment of this problem have been made using the Atomistic Green's Function (AGF) method to analyze how phonons are transmitted and reflected by defects \cite{ZYOng:PRB15_Efficient,ZYOng:PRB18_Atomistic,ZYOng:JAP18_Tutorial}. For instance, using a plane stress quasi-one-dimensional FEM model, modal transmission has been calculated recently for a finite phononic crystal \cite{gu2019simulation}. In this paper, we go beyond that and not only solve the scattering problem for a 1-D finite phononic crystal using AGF, but also show how to apply the formalism to waveguide scattering problems. As we shall see later, the close conceptual analogy between the phonons and acoustic/elastic waves suggests that key insights from the AGF method can be transposed to the more general wave scattering problem.\par

In what follows, we first review the concepts of open systems and the Atomistic Green's Function method (AGF) in sections (\ref{sec:os}) and (\ref{sec:agf}) and then elaborate on the implementation of the latter to solve scattering problems in a 1D elastic wave problem. Section (\ref{sec:waveguideScatter}) moves one step further and discusses the implementation of the AGF for solving a 2D in-plane scattering problem in an elastic waveguide. There, the discretization using FEniCS, finding the dispersion relations of Lamb waves using AGF, solving the scattering problem, and some numerical examples are discussed. The machinery of the Decimation technique, which is an iterative method, is also reviewed in the appendix to be used in finding the surface Green's function matrices of the 2D problem.

\section{Open systems}\label{sec:os}

\begin{figure}[htp]
\centering
\includegraphics[scale=.35]{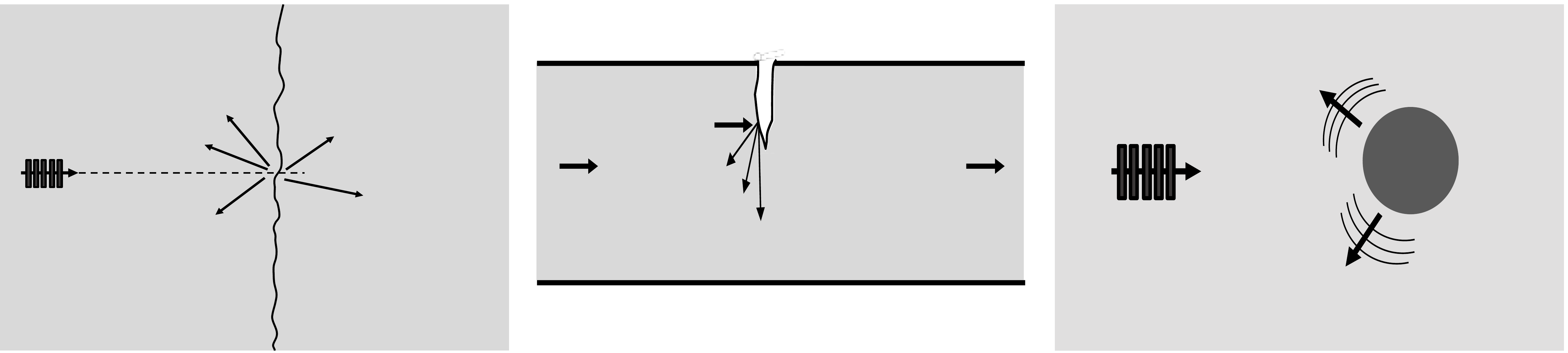}
\caption{Some scattering problems of interest. From left to right, the figures show scattering at a rough interface between two half planes, scattering from a notch in a waveguide, and acoustic scattering from a finite scatterer. }
\label{opensys}
\end{figure}

Of interest in this paper are scattering problems involving acoustic and elastic waves. Some schematics of such problems are shown in Fig. (\ref{opensys}). The scattering problems consist of a finite scatterer -- termed a device -- embedded in an infinite media -- termed the environment. This view of the scattering problem is termed an open system since the energy contained in the device is allowed to dissipate away to infinity \cite{KHODAVIRDI2022104399}. In this problem, admissible waves traveling in the environment get scattered by the device. These admissible waves are solutions of the dispersion relation which characterizes wave propagation in the environment (with the device removed). The information about the scattered waves is encapsulated in the scattering matrices (reflection matrix $\bm{r}$ and transmission matrix $\bm{t}$). The wave scattering dynamics of the entire infinite problem (environment+device) is characterized by an infinite-dimensional Hamiltonian $\bm{H}$ which, in general, cannot be calculated explicitly. The dynamics of the device, isolated from the environment, is characterized by a finite dimensional Hamiltonian $\bm{H}_D$ which, by itself, cannot provide information about the scattering processes which happen in the infinite problem. The purpose of the formalism presented in this paper is to show how to combine $\bm{H},\bm{H}_D$ into an $\bm{H}_{eff}$ which has the same dimensionality as $\bm{H}_D$ but \textit{does} provide the full information about the scattering processes. $\bm{H}_{eff}$ is called the effective Hamiltonian of the device and, once calculated, can be used to calculate the Green's function of the device, $\bm{G}_D$, through the inverse relationship:
\begin{align*}
\bm{G}_D=\left[\omega^2-\bm{H}_{eff}\right]^{-1}
\end{align*}
$\bm{G}_D$ can then be used to calculate $\bm{r},\bm{t}$ and other derived scattering metrics through straightforward relationships elucidated later in the paper. The calculation of $\bm{H}_{eff}$ is achieved through the formalism of the Atomistic Green's Function (AGF) technique and the main conceptual idea is the reduction of the infinite-dimensional $\bm{H}$ to a finite-dimensional $\bm{H}_{eff}$ by invoking the translational symmetry of the environment and by employing a numerical technique called \textit{decimation}.

\section{The Atomistic Green's Function method}\label{sec:agf}

The Atomistic Green's Function (AGF) method has its origin in a numerical technique used in nanoscale thermal transport research where the understanding of phonon transport in semiconductors and insulators is a longstanding challenge critical to the efficient thermal management of nanoscale electronic devices. In semiconductors and insulators, heat conduction is primarily mediated by phonons, which are quantized excitations of the wavelike normal modes, with each phonon carrying a (pseudo)momentum of $q$  and energy of $\omega$ as it propagates. In a real crystal lattice, phonons are scattered by local defects and boundaries, resulting in momentum dissipation and resistance to heat conduction. Hence, an accurate treatment of phonon scattering is needed for modeling the physical processes that affect solid state heat conduction.

In recent years, considerable effort has been expended by the nanoscale thermal transport community in the development of numerical techniques to describe elastic phonon scattering. In particular, the AGF method, pioneered by Mingo and Yang~\cite{NMingo:PRB03_Phonon}, has proved to be a powerful computational tool for modeling coherent phonon transmission and heat conduction in low-dimensional nanostructures, such as silicon nanowires and molecular junctions, because it derives the quantum-mechanical lattice heat flux from the relatively simple classical atomic equations of motion. In addition, the atomistic fidelity of the AGF method allows us to understand how the atomistic structure of defects affect to thermal transport, because its inputs are the interatomic force constants (IFCs) obtained from ab-initio or empirical models. 

Conceptually, the AGF method can be rigorously derived from the theory of nonequilibrium Green's functions (NEGF) for phonons~\cite{JSWang:EPJB08_Quantum, JSWang:FrontPhys14_NEGF} and its numerical implementation draws on well-established techniques and algorithms~\cite{WZhang:JHT07_Simulation, WZhang:NHT07_Atomistic} developed for studying ballistic electron transport in \emph{open} quantum systems. In the typical AGF setup, the system is partitioned into three components: the finite \emph{device}, in which the scattering takes place, and the semi-infinite left and right \emph{leads} that sandwich the device. The heat flux in the device comprises partially transmitted phonons originating from one lead and propagating towards the other. A highly attractive feature of the AGF method is that in the harmonic limit, this heat flux is computed exactly from the frequency-domain Green's function matrix of the device which has a finite subset of the total number of degrees of freedom. Unlike more traditional approaches such as wave packet simulations, there is no attempt to model the atomic displacements in the device and leads. Instead, the AGF method is considerably more computationally efficient because it uses primarily the IFCs of the device and the infinite degrees of freedom associated with the leads are absorbed into the device Green's function via the so-called self-energy terms.  

Nonetheless, a drawback of the traditional AGF method is its inability to describe mode-resolved phonon transmission in terms of the bulk phonon dispersion that characterizes the modes of the leads. To remedy this, Ong and Zhang~\cite{ZYOng:PRB15_Efficient} developed a computationally efficient extension of the AGF method, connecting mode-resolved phonon transmission to polarization, frequency and momentum. In a following paper, Ong~\cite{ZYOng:PRB18_Atomistic} derived the forward and backward scattering matrix amplitudes that describe transmission and reflection. A key idea in the extended AGF method is the frequency-dependent Bloch matrix~\cite{TAndo:PRB91_Quantum, ZYOng:JAP18_Tutorial,PAKhomyakov:PRB05_Conductance} which is derived from the surface Green's function of the lead and associated with its translational symmetry. In condensed matter physics, researchers have taken advantage of the modal resolution of the extended AGF method to study phonon Anderson localization~\cite{RHu:PRB21_Direct}, the specularity of phonon-boundary scattering~\cite{ZYOng:PRB20_Structure, ZYOng:EPL21_Specular, QSong:PRB21_Evaluation}, phonon transmission through amorphous silicon~\cite{LYang:PRB18_Phonon}, and valley filtering of phonons~\cite{XChen:PRB19_Valley}.

Although the AGF method has hitherto been used predominantly in the research of phonon scattering in condensed matter physics, there are grounds to believe that the its utility may go beyond physics problems. Firstly, the calculation of transmission and reflection coefficients for individual modes is important in both nanoscale thermal transport and elastic wave research. Secondly, the atomic equations of motion in the lattice are second order in time like the wave equations commonly found in acoustics and elastodynamics. Thirdly, a discrete translational symmetry analogous to that of a crystal lattice is introduced when space is discretized into a uniform grid. The resultant linear equations for the field variables in  acoustics and elastodynamics bear a strong mathematical similarity to the atomic equations of motion for the crystal lattice. This suggests that the insights and techniques from the extended AGF method can be brought to bear on the problem of wave transmission and reflection in these fields. In the following example of the 1D elastic wave, we introduce elements of the AGF method in an elastodynamic context and show how the AGF method is used to calculate transmission coefficients.

\subsection{1D elastic wave}

\begin{figure}[htp]
\centering
\includegraphics[scale=.34]{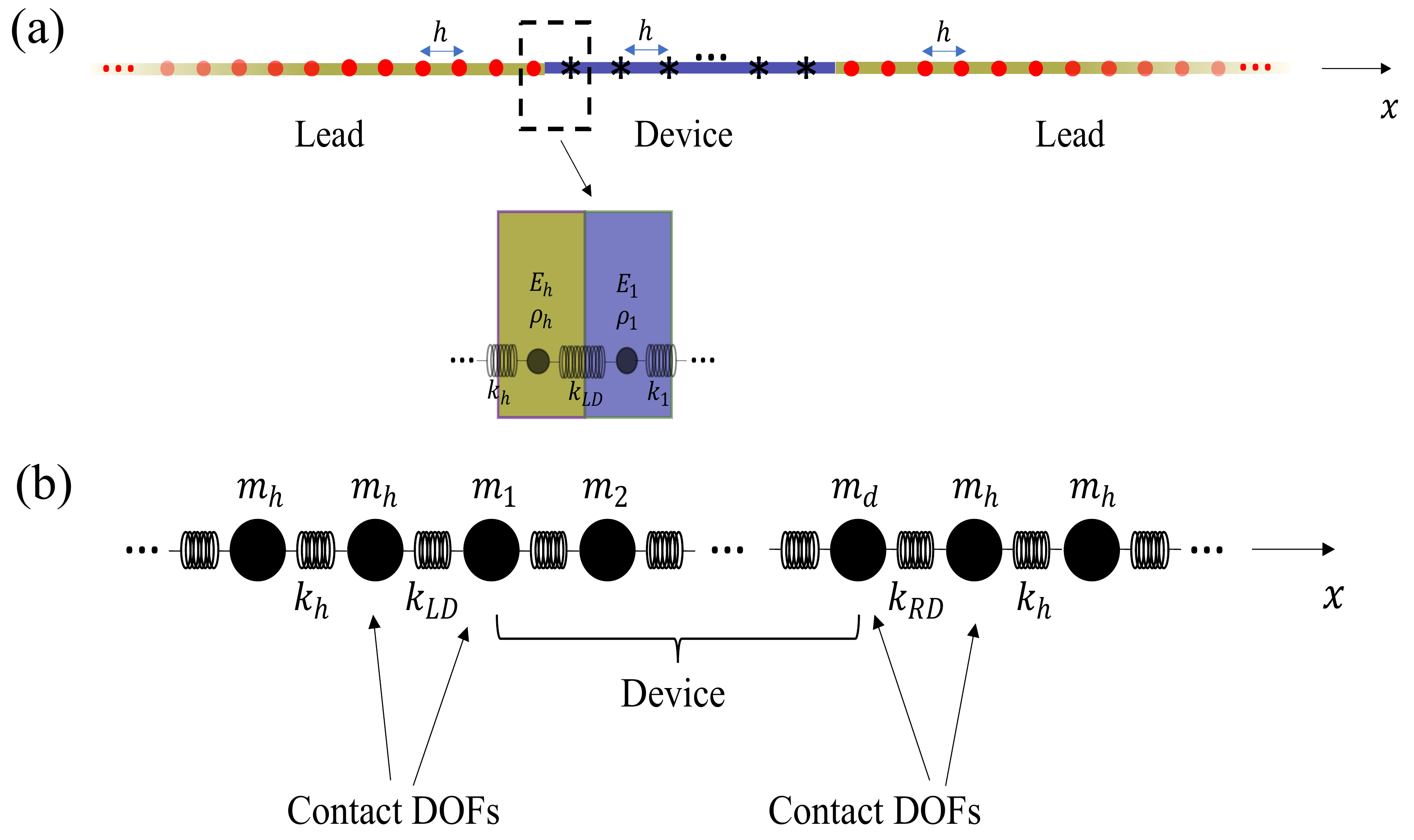}
\caption{(a) Discretized version of the three parts (lead-device-lead) of a scattering problem involving homogeneous mediums, (b) the same problem modeled with masses and springs whose values are derived from the descretization above.}
\label{fig:1D}
\end{figure}

We begin the discussion of the AGF method by considering the case of 1D elastic wave scattering. The important material properties for this class of problem are the Young's modulus $E(x)$ and the density $\rho(x)$. Consider the problem domain shown in \fref{fig:1D} which consists of a finite region of heterogeneity (device $D$) coupled to two semi-infinite homogeneous regions (leads $L,R$) on either side. The frequency-domain governing wave equation for this problem is:
\begin{equation}\label{eq:elasti1D}
-\frac{\partial}{\partial x}\left(E(x)\frac{\partial u}{\partial x}\right)=\omega^2\rho(x)u;\quad x\in D,L,R
\end{equation}
where $u(x)$ is the axial deformation and $\omega$ denotes the angular frequency. In addition, we have stress continuity at the two interfaces. One can convert this continuous problem into a discrete one by discretizing the space $x$ using a uniform grid $x_i$ with spacing $h$ and employing, as an example, finite difference (FD) schemes to approximate the space derivatives. The device is discretized into $n_D$ degrees of freedom (dof) whereas the leads are discretized into $N$ dof each where $N$ goes to infinity. The 1-D problem can also be discretized by considering the medium as an infinite chain of springs and masses. We take the bulk medium (leads) to have the homogeneous material properties $E_h,\rho_h$. By discretizing the medium with spatial interval $h$, we can easily determine the equivalent mass in the leads as $m_h=h\rho_h$ and the equivalent spring constant as $k_h=E_h/h$. Inside the device, the medium can be heterogeneous and, therefore, we talk about discrete masses $m_i=h\rho_i,i=1,...n_D$. The masses $m_1,m_D$ are connected to the leads through contact spring constants $k_{LD},k_{RD}$ respectively. These are given by $k_{LD}=2E_hE_1/h(E_h+E_1)$ and $k_{RD}=2E_hE_D/h(E_h+E_D)$. 

The time-reduced equations for the mass-spring problem can be framed in the following form now:
\begin{equation}
    \bm{Ku}=\omega^2 \bm{Mu}
\end{equation}
where the matrices are infinite dimensional as they correspond to the discretization of the infinite extent problem. The matrices $\bm{K}$ and $\bm{M}$ denote the tridiagonal stiffness and diagonal mass matrix, respectively. The above can be transformed into the canonical form used in the AGF method:
\begin{equation}\label{eq:AGFHarmonic}
    (\omega^2 \bm{I}-\bm{H})\bm{u}=0
\end{equation}
through the introduction of the harmonic matrix $\bm{H}$ with elements $H_{ij}=(M_iM_j)^{-1/2}K_{ij}$ where $M_i$ denotes the $i$-th diagonal element of $\bm{M}$. Due to the only-local effect of interactions in the lattice, $\bm{H}$, which is a  real symmetric matrix, has a banded structure of the form:

\begin{frame}
\footnotesize
\setlength{\arraycolsep}{5pt} 
\medmuskip = 7mu 
\newcommand{\bigzero}{\mbox{\normalfont\Large\bfseries 0}}
\newcommand{\rvline}{\hspace*{-\arraycolsep}\vline\hspace*{-\arraycolsep}}
\begin{equation}
\bm{H}=
\begin{pmatrix}
  \begin{matrix}
  \ddots & \ddots &&\vdots\\
 \ddots&&&\bm{H}_L^{01}\\
 &&&\\
\dots&0&\bm{H}_L^{10}&\bm{H}_L^{00}
  \end{matrix}
  & \rvline &
 \begin{matrix}
&&&\\
&&&\\
&&\bm{H}_{LD}&\\
&&&\\
&&&
\end{matrix}& \rvline &\bigzero\\
\hline \begin{matrix}
&&&\\
&&&\\
&&\bm{H}_{DL}&\\
&&&\\
&&&
\end{matrix}& \rvline &
  \begin{matrix}
  \frac{k_h+k_{LD}}{h\rho_h}&-\frac{k_{LD}}{h\sqrt{\rho_h\rho_{1}}} &0&\dots\\
 -\frac{k_{LD}}{h\sqrt{\rho_h\rho_{1}}}&\ddots&&\vdots\\
 \vdots&&\ddots&-\frac{k_{RD}}{h\sqrt{\rho_h\rho_{n_D}}}\\
 \dots&0&-\frac{k_{RD}}{h\sqrt{\rho_h\rho_{n_D}}}&\frac{k_{RD}+k_h}{h\rho_h}
\end{matrix}&\rvline&\begin{matrix}
&&&\\
&&&\\
&&\bm{H}_{DR}&\\
&&&\\
&&&
  \end{matrix}\\ \hline \bigzero & \rvline &
  \begin{matrix}
&&&&\\
&&&\\
&&\bm{H}_{RD}&&\\
&&&&\\
&&&&
  \end{matrix}&\rvline&\begin{matrix}
\bm{H}_{R}^{00}&\bm{H}_{R}^{01}&0&\dots\\
 &&&\\
 \bm{H}_{R}^{01}&&&\ddots\\
 \vdots&&\ddots&\ddots
  \end{matrix}
\end{pmatrix}
\label{eq:HBandedExp}
\end{equation}
\end{frame}

Here, $\bm{H}_L^{00}=\bm{H}_R^{00}=\frac{2k_h}{h\rho_h}$ and $\bm{H}_L^{10}=\bm{H}_R^{01}=\frac{-k_h}{h\rho_h}$. $\bm{H}_{LD},\bm{H}_{RD}$ also represent the coupling matrices which couple the leads to the device. For 1D problems with only local interactions, all the elements of the coupling matrices are zero except $\bm{H}_{LD}(N,1)=\bm{H}_{DL}(1,N)=\frac{-k_h}{h\rho_h}$ and, $\bm{H}_{RD}(1,n_D+2)=\bm{H}_{DR}(n_D+2,1)=\frac{-k_h}{h\rho_h}$. $\bm{H}$ can be written in a compact form:

\begin{equation}
\label{eq:Hbandedform3}
\bm{H}=
\begin{pmatrix}
\bm{H}_L & \bm{H}_{LD} & \bm{0}\\
\bm{H}_{DL} & \bm{H}_D & \bm{H}_{DR}\\
\bm{0} & \bm{H}_{RD} & \bm{H}_R
\end{pmatrix}
\end{equation}
The real symmetric structure of $\bm{H}$ implies that $\bm{H}_{LD}=\bm{H}_{DL}^\dagger,\bm{H}_{RD}=\bm{H}_{DR}^\dagger$. $\bm{H}_{D}$ is a square matrix of size $(n_D+2,n_D+2)$ whereas $\bm{H}_{LD},\bm{H}_{RD}$ are matrices of size $(N,n_D+2)$. Here, we highlight  that the matrix in Eq. (\ref{eq:HBandedExp}) is assembled by assuming that the immediate degrees of freedom of both the left and right leads are part of the device. This simplifies later calculations. The main goal now is to extract from the harmonic matrix above, the frequency-dependent Green's function of the system. The frequency $\omega$ is set equal to that for the incoming and outgoing waves. The Green's function matrix is formally given by:
\begin{equation}\label{eq:totalGF}
\bm{G}=\left[(\omega+i\eta)^2-\bm{H}\right]^{-1}
\end{equation}
where $0<\eta\ll \omega$ and $i\eta$ is added to account for causality. Since the size of $\bm{H}$ is infinite, $\bm{G}$ cannot be computed using the above in a straightforward fashion. Of greater importance than the full Green's function matrix is the device subset of the matrix, $\bm{G}_D$ which is finite:
\begin{equation}\label{eq:deviceGF}
\bm{G}_D=\left[(\omega+i\eta)^2-\bm{H}_D-\bm{\Sigma}_1-\bm{\Sigma}_2\right]^{-1}
\end{equation}
where $\bm{\Sigma}_1,\bm{\Sigma}_2$ are the frequency-dependent self-energy matrices corresponding to the left and right contacts respectively. These matrices are given by:
\begin{equation}\label{eq:selfEnergies}
\bm{\Sigma}_1=\bm{H}_{DL}\bm{g}_L\bm{H}_{DL}^\dagger;\quad \bm{\Sigma}_2=\bm{H}_{DR}\bm{g}_R\bm{H}_{DR}^\dagger
\end{equation}
In the above, $\bm{g}_L,\bm{g}_R$ are the Green's functions of the left and right semi-infinite leads when they are uncoupled from the device:
\begin{equation}\label{eq:uncoupledGF}
\bm{g}_L=\left[(\omega+i\eta)^2-\bm{H}_L\right]^{-1}; \quad \bm{g}_R=\left[(\omega+i\eta)^2-\bm{H}_R\right]^{-1}
\end{equation}

We note that if the self-energy matrices could be calculated then we would have determined the finite effective harmonic matrix (effective Hamiltonian) of the device as well:
\begin{equation}\label{eq:effectiveHamiltonian}
\bm{H}_{eff}(\omega)=\bm{H}_D+\bm{\Sigma}_1+\bm{\Sigma}_2
\end{equation}
Even though the self-energy matrices are of finite size $(n_D+2,n_D+2)$, their calculation involves the infinite matrices $\bm{H}_{DL},\bm{H}_{DR}$ as well as the infinite Green's function matrices of the uncoupled left and right leads $\bm{g}_L,\bm{g}_R$. However, since the matrices $\bm{H}_{DL},\bm{H}_{DR}$ have a single non-zero element each, we can exploit this fact to calculate the self-energy matrices. To be more explicit, since the only non-zero element of $\bm{H}_{DL}$ is $\bm{H}_{DL}(1,N)$ and the only non-zero element of $\bm{H}_{DL}^\dagger$ is $\bm{H}_{DL}^\dagger(N,1)$, the only element of $\bm{g}_L$ of interest and consequence is $\bm{g}_L(N,N)$. Similarly the only element of $\bm{g}_R$ of interest and consequence is $\bm{g}_R(1,1)$. $\bm{g}_L(N,N)$ and $\bm{g}_R(1,1)$ correspond to the lead Green's function components at their surfaces where the leads are connected to the device. They are, therefore, called surface Green's functions and their calculation is an important aspect of the method under discussion. Due to their importance in later equations, we term these $\bm{g}_{L}^\text{surf} \equiv \bm{g}_{L}(N,N)$, $\bm{g}_{R}^\text{surf} \equiv \bm{g}_{R}(1,1)$, $\bm{H}_{DL}^\text{surf} \equiv \bm{H}_{DL}(1,N)$, and $\bm{H}_{DR}^\text{surf} \equiv \bm{H}_{DR}(N,1)$. For the current problem, the surface Green's functions can be computed analytically.

\subsubsection{Analytical computation of surface Green's functions}

We note that since the only non-zero term in $\bm{H}_{DL}$ is $\bm{H}_{DL}(N,1)$, $\bm{\Sigma}_1=\bm{H}_{DL}\bm{g}_L\bm{H}_{DL}^\dagger$ is zero everywhere except for the element $\bm{\Sigma}_1(1,1)$ and this term is equal to $\bm{H}_{DL}^\text{surf} \bm{g}_{L}^\text{surf} (\bm{H}_{DL}^\text{surf})^\dagger$. Similarly, $\bm{\Sigma}_2$ is zero everywhere except $\bm{\Sigma}_2(n_D+2,n_D+2)=\bm{H}_{DR}^\text{surf} \bm{g}_{R}^\text{surf} (\bm{H}_{DR}^\text{surf})^\dagger$. Since we have $\bm{g}_L=\left[(\omega+i\eta)^2-\bm{H}_L\right]^{-1}$ and  $\bm{g}_L(N,N)$ is of special significance, we can partition the Green's function of the left lead as:
\begin{frame}
\footnotesize
\setlength{\arraycolsep}{5pt} 
\medmuskip = 7mu 
\newcommand{\bigzero}{\mbox{\normalfont\Large\bfseries 0}}
\newcommand{\rvline}{\hspace*{-\arraycolsep}\vline\hspace*{-\arraycolsep}}
\begin{equation}
    \label{GL2}
    \bm{g}_L=\begin{pmatrix}
\begin{matrix}
&\left(\omega+i\eta\right)^2\bm{I}-\bm{H}_{L}&
\end{matrix} & \rvline &\begin{matrix}
0\\
\vdots\\
0\\
\bm{H}_{L}^{01}
\end{matrix} \\ \hline \begin{matrix}
0&\dots&0&\bm{H}_{L}^{10}
\end{matrix} & \rvline & \begin{matrix}
&\left(\omega+i\eta\right)^2\bm{I}-\bm{H}_{L}^{00}&
\end{matrix}
\end{pmatrix}^{-1}
\end{equation}
\end{frame}
For the current 1-D problem, the blocks $\bm{H}_L^{10}$ and $\bm{H}_L^{01}$ are $1\times1$, so the quarters in the top right and bottom left are column and row vectors respectively, with only one non-zero element. Using established matrix identity, we can write the $\bm{g}_L(N,N)$ term as:
\begin{equation}
\label{GL4}
   \bm{g}_L\left(N,N\right)= \left\{\left(\omega+i\eta\right)^2-\bm{H}_{L}^{00}-\bm{Y}\left[\left(\omega+i\eta\right)^2\bm{I}-\bm{H}_{L}\right]^{-1}\bm{Y}^\dagger\right\}^{-1}
\end{equation}
where $\bm{Y}=\left(0\quad...\quad \bm{H}_L^{10}\right)$. The expression $\left[\left(\omega+i\eta\right)^2\bm{I}-\bm{H}_{L}\right]^{-1}$ is nothing but $\bm{g}_L$. The product $\bm{Y}\bm{g}_L\bm{Y}^\dagger$ is simply equal to $\bm{H}^{10}_L \bm{g}_L(N,N) \bm{H}^{01}_L$. We, therefore, have the relation:

\begin{equation}
    \bm{g}_L(N,N) = \left[(\omega+i\eta)^2 - \bm{H}^{00}_L - \bm{H}^{10}_L \bm{g}_L(N,N) \bm{H}^{01}_L\right]^{-1}
\end{equation}
Substituting the parameters specific to the present problem:
\begin{equation}
    \bm{g}_L\left(N,N\right)=\left[\left(\omega+i\eta\right)^2-\frac{2k_h}{h\rho_h}-\frac{k_h}{h\rho_h} \bm{g}_L\left(N,N\right)\frac{k_h}{h\rho_h}\right]^{-1}
\end{equation}
We obtain an equation that is quadratic in $\bm{g}_L(N,N)$ and yields two possible solutions of $\bm{g}_L(N,N)$ from its roots. The correct solution is obtained by considering the weak coupling limit. In the weak coupling ($k_h \rightarrow 0$) limit, we expect the surface of the lead to behave asymptotically as a decoupled independent oscillator such that $\lim_{k_h \rightarrow 0}g_L(N,N)\sim 1/\omega^2$. This results in the following admissible solution for $\bm{g}_L(N,N)$:
\begin{equation}
\label{surfaceGreensL}
    \bm{g}_L\left(N,N\right)=\frac{\left[\left(\omega+i\eta\right)^2-\frac{2k_h}{h\rho_h}\right]-\sqrt{\left[\left(\omega+i\eta\right)^2-\frac{2k_h}{h\rho_h}\right]^2-4\left(\frac{k_h}{h\rho_h}\right)^2}}{2\left(\frac{k_h}{h\rho_h}\right)^2}
\end{equation}
which has the correct asymptotic behavior in the weak coupling limit. 
Proceeding similarly for the right lead, we have the surface Green's function $\bm{g}_R(1,1)=\bm{g}_L(N,N)$. Finally, the only non-zero term of the self-energy matrices can also be calculated:
\begin{equation}
    \label{eq:SelfEng121D}
    \bm{\Sigma}_1\left(1,1\right)=\left(\frac{-k_h}{h\rho_h}\right)^2\bm{g}_L(N,N); \quad \bm{\Sigma}_2\left(n_D+2,n_D+2\right)=\left(\frac{-k_h}{h\rho_h}\right)^2\bm{g}_R(1,1)
\end{equation}

\subsubsection{Transmittance}
Once the device Green's function, $\bm{G}_D$, has been determined, one may use it to calculate various scattering properties of the system. If there is an incident wave in the left lead with unit amplitude: $\exp(ikx)$ with $k=\omega\sqrt{\rho/E}$, it gives rise to a transmitted wave in the right lead: $T(\omega)\exp(ikx)$. The function $T(\omega)$ is called the transmission coefficient and $T^2(\omega)$ is the total transmitted energy, also called the transmittance. The transmittance can be directly calculated from the matrices already calculated through the use of the Caroli formula\cite{li2012generalized}:
\begin{align}\label{eq:transmittance}
\nonumber \bm{A}_1=i\left[\bm{g}_L-\bm{g}_L^\dagger\right];\quad \bm{A}_2=i\left[\bm{g}_R-\bm{g}_R^\dagger\right]\\
\nonumber \bm{\Gamma}_1=\bm{H}_{LD}\bm{A}_1\bm{H}_{LD}^\dagger;\quad \bm{\Gamma}_2=\bm{H}_{RD}\bm{A}_2\bm{H}_{RD}^\dagger\\
T^2(\omega)=\mathrm{Tr}\left[\Gamma_1\bm{G}_D\Gamma_2\bm{G}_D^\dagger\right]
\end{align}

In addition to the Caroli expression, for the current 1D problem, the transmission can be found through other methods and comparisons can be made. For instance, consider a simple problem where the device itself is homogeneous with material properties $E_0,\rho_0$ and length $L$. We assume that the lead-device interfaces are at $x=-L/2,L/2$. We assume that there exists an incoming wave in the lead of unit amplitude $e^{ik_hx}$ where $k_h=\omega/\sqrt{E_h/\rho_h}$ leading to a reflected wave in the left lead $Re^{-ik_hx}$ and a transmitted wave in the right lead $Te^{ik_hx}$. The displacement field in the device is made up of the waves which are admitted in the region: $Ae^{ik_0h}+Be^{-ik_0h}$ where $k_0=\omega/\sqrt{E_0/\rho_0}$. The four unknowns in the problem $R,T,A,B$ are now solved by imposing displacement and stress continuity relations at $x=-L/2,L/2$, allowing us to calculate $T(\omega)$. This is a standard technique which we call the mode-matching method.
\begin{figure}[htp]
\centering
\includegraphics[scale=0.25]{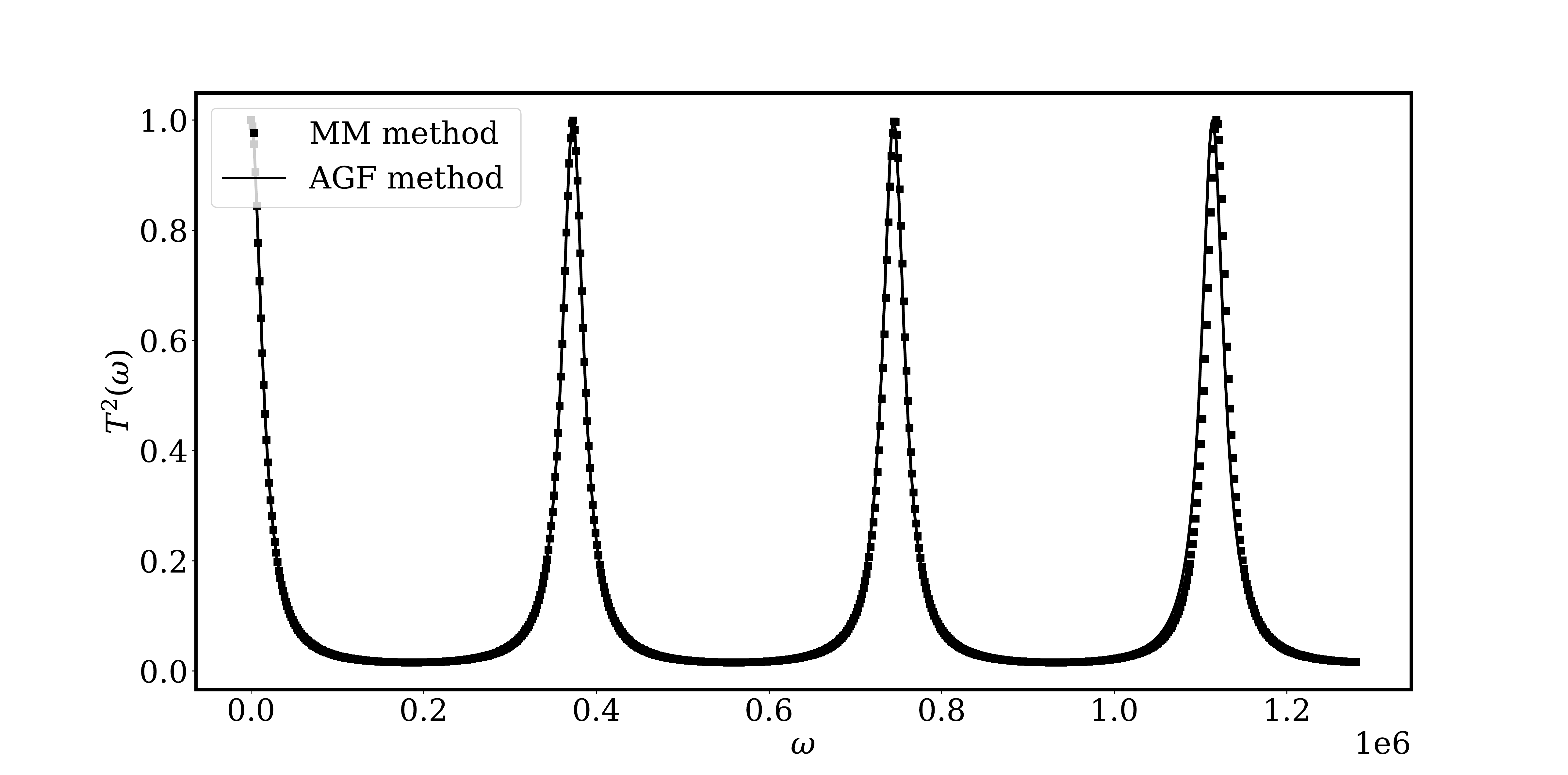}
\caption{Transmission vs. angular frequency for a 1D scattering problem where the scatterer is a region of material heterogeneity.}
\label{Numerical1D}
\end{figure}
\fref{Numerical1D} shows the results where we compare the transmittance calculated from the AGF formulation with that calculated from the mode matching method described above. In creating this figure, we have considered the following material and geometric properties: $E_h=8$ GPa, $\rho_h=1180$ kg/m$^3$, $E_D=300$ GPa, $\rho_D=8000$ kg/m$^3$, $L=0.051$ m, and $A=1$ m$^2$. The lattice constant used for discretization in the AGF method is $h=0.0017$ m. The results show excellent agreement between the AGF and mode matching results. The results diverge slightly at the high end of the frequency range which is expected since the AGF formalism solves a discretized version of a continuous problem whereas the mode matching method solves the continuous problem directly.

\begin{figure}[htp]
\centering
\includegraphics[scale=0.45]{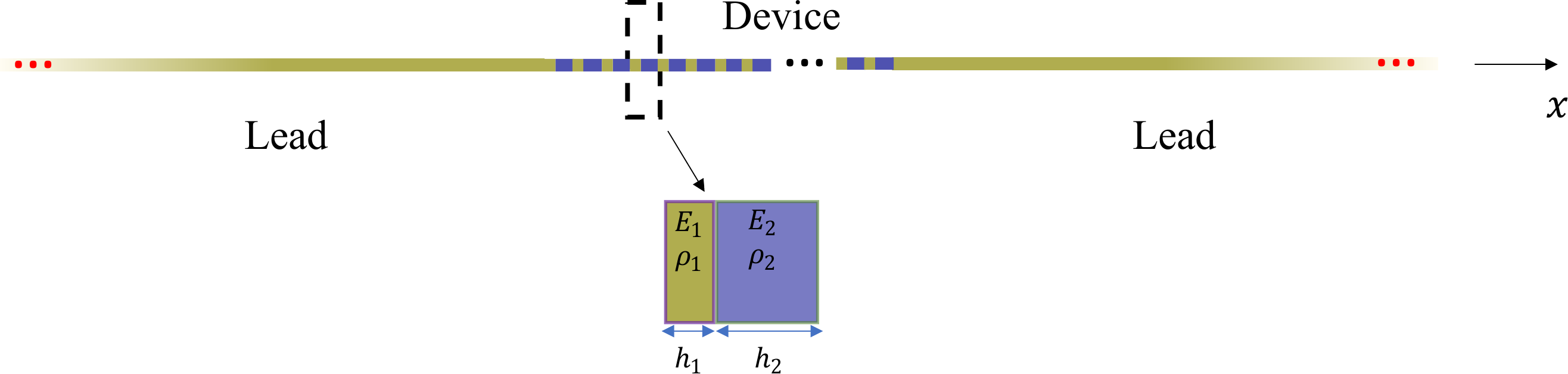}
\caption{Schematic of a 1D finite phononic crystal connected to two semi-infinite 1D leads.}
\label{phononic1D1schematic}
\end{figure}
A further example may be considered where the device is a finite phononic crystal (as shown in \fref{phononic1D1schematic}). The phononic crystal is made up of a bi-phase unit cell with the following material and geometric properties: $E_1=8$ GPa, $\rho_1=1180$ kg/m$^3$, $h_1=2\times10^{-3}$ m, $E_2=300$ GPa, $\rho_2=8000$ kg/m$^3$, $h_2=1.62\times10^{-4}$ m, where $h_1$ and $h_2$ are the width in phase one and two of the unit cell, respectively. The device is made up of 10 such unit cells. This new configuration makes little difference to the AGF method but the mode matching method for this case is more involved. However, the details are standard and given in Ref.~\cite{srivastava2014limit}. 
\begin{figure}[htp]
\centering
\includegraphics[scale=0.183]{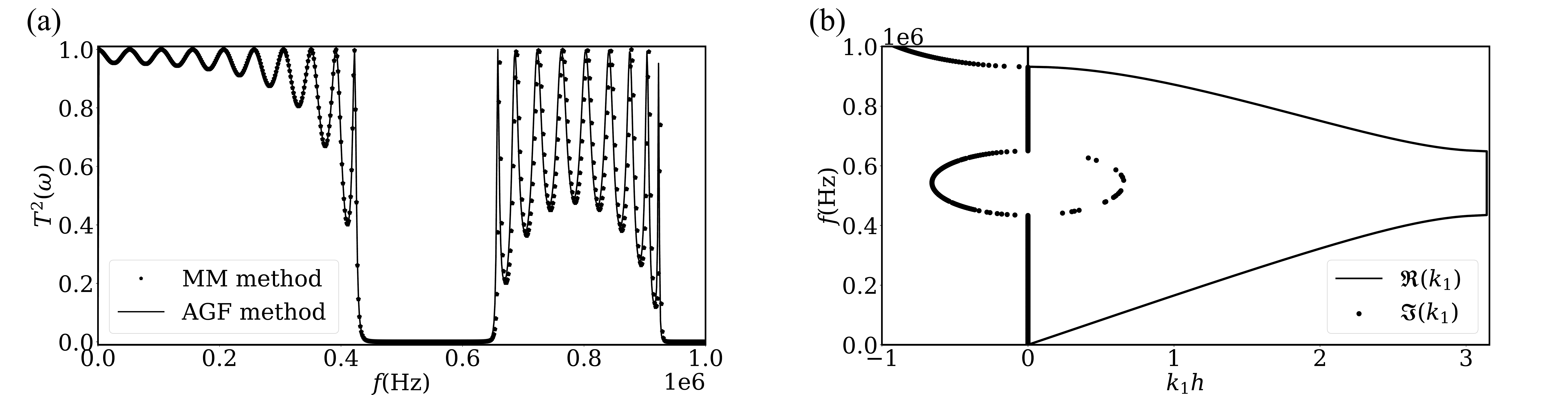}
\caption{ Scattering results for scattering from a 1D phononic crystal. (a) Transmission vs. frequency calculations using the Atomistic Green's Function method and the Mode Matching method. (b) Bandstructure calculations using the Transfer Matrix Method.}
\label{transmissionandBS}
\end{figure}
\fref{transmissionandBS}a shows a comparison between the AGF method and the mode-matching method for the calculation of the transmittance. First, we note that the AGF method provides a near-perfect match with the results of the mode-matching method over the frequency range considered. The phononic crystal exhibits a bandgap in the frequency range, $\left[433,650\right]$ kHz, and the AGF calculations are able to capture this phenomenon as the calculated transmittance values go to zero in that frequency band.

\section{Waveguide problem}\label{sec:waveguideScatter}

Next, we consider the application of the AGF method to the scattering problem in a waveguide. This problem will serve as a platform to elucidate further advanced concepts in the theory and application of the AGF method. The problem involves interface coupling over several degrees of freedom and, as a result, it is not possible to evaluate the surface Green's function in an analytical fashion here. We will demonstrate the technique of decimation which will allow us to calculate the surface Green's function here. The problem also involves the scattering of multiple wave-modes (Lamb waves), a full understanding of which requires the evaluation of the \textit{S}-matrix. We will show how the \textit{S}-matrix automatically emerges from the matrices which are already calculated as part of the AGF method. 
\begin{figure}[htp]
\centering
\includegraphics[scale=0.45]{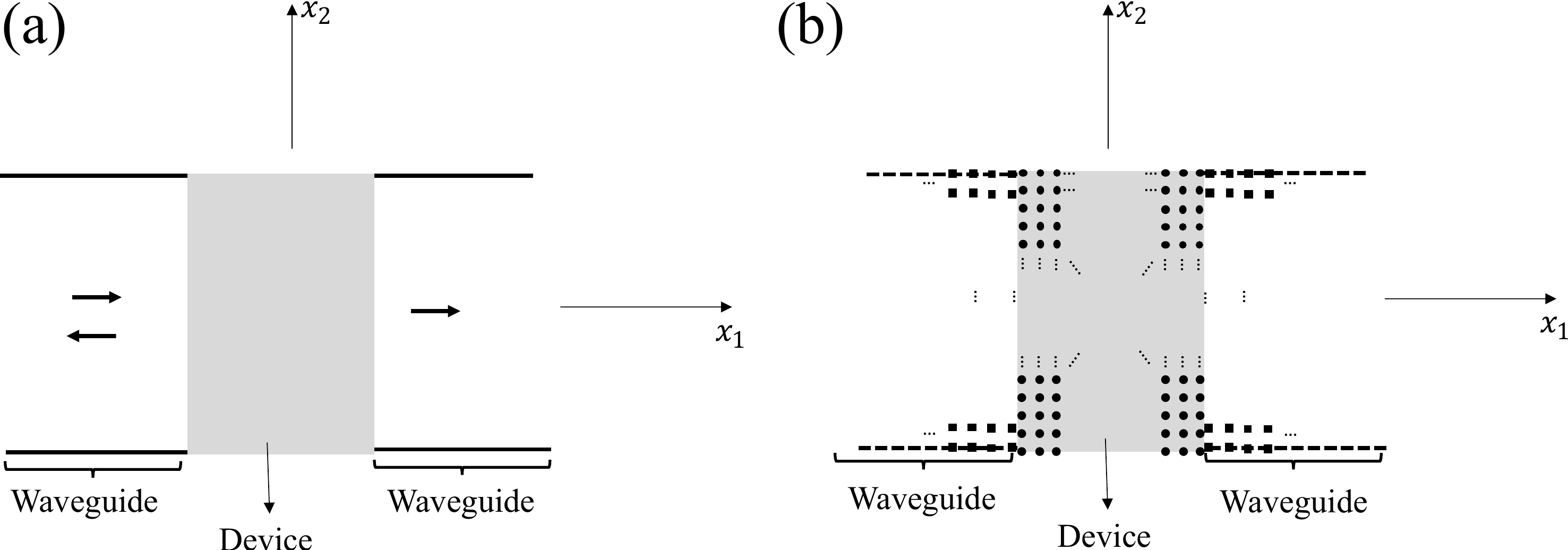}
\caption{(a) Schematic of scattering in a 2D waveguide. (b) Schematic showing a discretized version of the waveguide problem.}
\label{Waveguide}
\end{figure}
\fref{Waveguide}a shows the schematic of the waveguide problem which involves a central scattering region (device, $D$) connected to two semi-infinite waveguides (leads). The field variables of interest in this problem are the two components of deformation $u_1,u_2$ and the relevant equations of motion (plane strain) are:
\begin{align}\label{eq:EOMWG}
\nonumber\rho\frac{\partial^2u_1}{\partial t^2}=\frac{\partial}{\partial x_1}\left[\left(2\mu+\lambda\right)\frac{\partial u_1}{\partial x_1}+\lambda\frac{\partial u_2}{\partial x_2}\right]+\frac{\partial}{\partial x_2}\left[\mu\left(\frac{\partial u_1}{\partial x_2}+\frac{\partial u_2}{\partial x_1}\right)\right]\\
\rho\frac{\partial^2u_2}{\partial t^2}=\frac{\partial}{\partial x_2}\left[\left(2\mu+\lambda\right)\frac{\partial u_2}{\partial x_2}+\lambda\frac{\partial u_1}{\partial x_1}\right]+\frac{\partial}{\partial x_1}\left[\mu\left(\frac{\partial u_1}{\partial x_2}+\frac{\partial u_2}{\partial x_1}\right)\right]
\end{align}
Here, $\lambda$ and $\mu$ are the Lam\'e constants. \fref{Waveguide}b shows the same problem, now discretized into a grid. We have specified the planes which couple the leads to the device. The formal frequency-domain equation of motion for the system is:
\begin{equation}\label{eq:AGFHarmonicWaveguide}
    (\omega^2 \bm{I}-\bm{H})\bm{\phi}(\omega)=0
\end{equation}
where $\bm{\phi}$ constitutes all the degrees of freedom of the system. Properly organized, the $\bm{H}$ matrix has the following structure:

\begin{equation}
\bm{H}=
\begin{pmatrix}
\bm{H}_L & \bm{H}_{LD} & \bm{0}\\
\bm{H}_{DL} & \bm{H}_D & \bm{H}_{DR}\\
\bm{0} & \bm{H}_{RD} & \bm{H}_R
\end{pmatrix}
\label{eq:bandedHWaveguide}
\end{equation}

In Eq.~(\ref{eq:bandedHWaveguide}), $\bm{H}_D$ is the harmonic submatrix corresponding to the device, and $\bm{H}_{LD},\bm{H}_{RD}$ are the harmonic submatrices corresponding to the interface region and the coupling between the device and leads. The left and right leads are sliced into principal layers of thickness $a_l,a_R$ perpendicular to the wave propagation direction, where $a_r=a_l=3\times10^{-3}$ m. These slices are enumerated $0,...\infty$ in the right lead and $-\infty,...0$ in the left lead. For the leads, the harmonic submatrices $\bm{H}_L^{00},\bm{H}_R^{00}$ correspond to the coupling of the degrees of freedom in each corresponding slice and the submatrices $\bm{H}_L^{01},\bm{H}_R^{01}$, which satisfy $(\bm{H}_L^{01})^\dagger=\bm{H}_L^{10}$, $(\bm{H}_R^{01})^\dagger=\bm{H}_R^{10}$, correspond to the coupling of each slice to the one on its right, such that 
\begin{equation*}
    \bm{H}_L = 
    \begin{pmatrix}
        \ddots & \ddots & \\
        \ddots & \bm{H}_L^{00} & \bm{H}_L^{01} \\
               & \bm{H}_L^{10} & \bm{H}_L^{00} 
    \end{pmatrix}
    \ , \ 
    \bm{H}_R = 
    \begin{pmatrix}        
        \bm{H}_R^{00} & \bm{H}_R^{01} & \\
        \bm{H}_R^{10} & \bm{H}_R^{00} & \ddots \\ 
                      & \ddots & \ddots  
    \end{pmatrix}    
    \ .
\end{equation*}
The principal layers in the leads are all assumed to be identical as a consequence of the translational symmetry of the environment.

\subsubsection{Calculating the submatrices - FEniCS}
In principle, there are several methods of calculating the submatrices in Eq. (\ref{eq:bandedHWaveguide}). Here we do so through the open-source finite element software FEniCS\cite{kirby2010,LoggEtal_11_2012,LoggEtal_10_2012}. The process involves considering the device and the leads together and meshing it over a regular grid. We note that, in general, automatic meshing in a finite element software will not result in a neat banded form of the $\bm{H}$ matrix as shown in Eq. (\ref{eq:bandedHWaveguide}). This is because the ordering of the degrees of freedom that results from the automatic mesh does not necessarily conform to the slice-based order inherent in the AGF method. Furthermore, the mass matrix which results from the finite element process is also generally not diagonal. To mitigate the first issue, we rearranged the degrees of freedom derived from automatic meshing to conform to the AGF scheme and reorganized the stiffness ($\bm{K}$) and mass ($\bm{M}$) matrices accordingly. To mitigate the second issue, we employed the lumped mass technique\cite{reddy2004introduction} which results in a diagonal mass matrix. 

\subsubsection{Wavemodes in the bulk waveguides}

The wavemodes in the leads can be directly evaluated from the submatrices $\bm{H}_L^{00},\bm{H}_R^{00}$ and $\bm{H}_L^{01},\bm{H}_R^{01}$. We focus on the left lead but the treatment in the right lead follows similarly. We introduce the Bloch factor $\lambda=\exp(ika_L)$ where $k$ is the wavenumber of the wave traveling along the waveguide in the left lead and $a_L$ denotes the interlayer spacing. Fixing $\omega$, we can find the admissible wavenumbers in the lead by solving the following quadratic eigenvalue problem\cite{ZYOng:PRB18_Atomistic}:
\begin{equation}\label{eq:bulkDispersion}
    -\bm{H}_{L}^{10}\bm{u}+\lambda(\omega^2\bm{I}_L-\bm{H}_{L}^{00})\bm{u}-\lambda^2\bm{H}_{L}^{01}\bm{u}=0
\end{equation}
We can determine the modes and their wavenumbers as a function of $\omega$ from the above equation. If the total number of dofs in each left slice is $N_L$ then there will be $2N_L$ solutions to the above eigenvalue equation. The modes can be classified as propagating and evanescent, and only propagating modes contribute to energy transfer along the waveguide. Half of the solutions will be rightward traveling (labeled with ``+'') and the other half will be leftward traveling (labeled with ``-''). The eigenvectors are similarly labeled with $\bm{u}_n(\pm)$ for $n=1,...N_L$. If we add a small imaginary part $i\eta$ to $\omega$ in Eq. (\ref{eq:bulkDispersion}) so that $\omega\rightarrow\omega+i\eta$, then the rightward traveling modes are the ones with $|\lambda|<1$, whereas the leftward traveling modes are the ones with $|\lambda|>1$. The rightward traveling modes can be further divided into propagating ($\lim_{\eta\rightarrow 0}|\lambda|=1$) and evanescent ($\lim_{\eta\rightarrow 0}|\lambda|<1$) states. Similarly, the leftward traveling modes are either propagating  ($\lim_{\eta\rightarrow 0}|\lambda|=1$) or evanescent ($\lim_{\eta\rightarrow 0}|\lambda|>1$).

\subsubsection{Device Green's function}
As in the 1D problem, we can calculate the device (effective) Green's function matrix from the following:
\begin{equation}\label{eq:deviceGFWaveguide}
\bm{G}_D=\left[(\omega+i\eta)^2-\bm{H}_D-\bm{\Sigma}_L-\bm{\Sigma}_R\right]^{-1}
\end{equation}
where the self-energy matrices are given by $\bm{\Sigma}_L=\bm{H}_{DL}\bm{g}_L^r\bm{H}_{DL}^\dagger, \bm{\Sigma}_R=\bm{H}_{DR}\bm{g}_R^r\bm{H}_{DR}^\dagger$
The retarded Green’s functions of the surface slice of the decoupled left and right lead, also called the surface Green’s functions, are given by the expressions:
\begin{align}\label{eq:surfaceGFWaveguide}
\nonumber \bm{g}_L^r=\left[(\omega+i\eta)^2-\bm{H}_L^{00}-\bm{\Sigma}_L^{00}\right]^{-1}\\
\bm{g}_R^r=\left[(\omega+i\eta)^2-\bm{H}_R^{00}-\bm{\Sigma}_R^{00}\right]^{-1}
\end{align}
where
\begin{align*}
\bm{\Sigma}_L^{00}=\bm{H}_{L}^{10}\bm{g}_L^r(\bm{H}_{L}^{10})^\dagger;\quad \bm{\Sigma}_R^{00}=\bm{H}_{R}^{01}\bm{g}_R^r(\bm{H}_{R}^{01})^\dagger
\end{align*}
The above equations are recursive in nature and represent the fact that the leads are translationally invariant. Unlike in the 1D case, the surface Green's functions cannot be calculated in an analytical form here. The surface Green's function needs to be solved either recursively or using the Decimation technique\cite{guinea1983effective}. Recursive solutions are slow to converge and, therefore, we use the Decimation technique which is described in Appendix (\ref{appdec}).

\subsubsection{Transmission and reflection matrices}
Once the device Green's function has been calculated, it can be used to calculate the transmission and scattering matrices. It is convenient to define some additional matrices here before dealing with the transmission and scattering matrices themselves. We first note that the wavemodes calculated for the bulk waveguides above correspond to the retarded Green's function solution satisfying the Sommerfeld radiation condition \cite{schot1992eighty}. We mark these wavemodes with modeshapes $\bm{u}_n^{r\pm}$ and corresponding eigenvalues $\lambda_n^{r\pm}$. We remember that there are $2N_L$ number of these modes for the left waveguide and $2N_R$ number of these modes for the right waveguide. There are additional solutions to the waveguide problem which do not satisfy the Sommerfield radiation condition and correspond to the advanced Green's function. These solutions can be obtained by employing the time reversal symmetric transformation for the waveguides: $t\rightarrow -t$ or by performing the substitution $\omega\rightarrow \omega-i\eta, \eta>0$. The solutions corresponding to the advanced Green's function will be represented by $\bm{u}_n^{a\pm}$ and corresponding eigenvalues $\lambda_n^{a\pm}$. We now define Bloch matrices $\bm{F}_L^{r+},\bm{F}_L^{r-}$ for the left lead which satisfy the linear eigenvalue equation:
\begin{align*}
\bm{F}_L^{r\pm}\bm{U}_L^{r\pm}=\bm{U}_L^{r\pm}\Lambda_L^{r\pm}
\end{align*}
where $\bm{U}_L^{r\pm}$ is a matrix where the columns consist of the normalized eigenvectors $\bm{u}_n^{r\pm}$ and $\Lambda_L^{r\pm}$ is a diagonal matrix where its diagonal elements are $\lambda_n^{r\pm}$. One can similarly define:
\begin{align*}
\bm{F}_L^{a\pm}\bm{U}_L^{a\pm}=\bm{U}_L^{a\pm}\Lambda_L^{a\pm}
\end{align*}
with similar matrices for the right waveguide. The Bloch matrices relevant for transmission matrix calculations can be evaluated from the surface Green's function matrix:

\begin{align*}
\left(\bm{F}_L^{a-}\right)^{-1}=\left[\bm{H}_L^{10}\bm{g}_L^a\right]^\dagger\\
\bm{F}_R^{r+}=\bm{g}_R^r\bm{H}_R^{10}
\end{align*}
where in $\bm{g}_R^r$ and $\bm{g}_L^a$ the superscript $r$ and $t$ representing retarded and advanced Green's matrices and they satisfy, $\bm{g}_R^r=(\bm{g}_L^a)^\dagger$. It is important to note that the Bloch matrices are not Hermitian, which can pose a problem when it comes to calculating scattering coefficients. Specifically, in cases where the eigenvectors have the same wave number $k$, and are wave number-degenerate, this can lead to inaccuracies. To overcome this issue, we employ a Gram-Schmidt procedure to orthonormalize the wave number-degenerate column eigenvectors \cite{leon2013gram}. Also necessary for the computation are the velocity matrices:
\begin{align}\label{eq:velocityMatrix}
\nonumber \bm{V}_L^+=\frac{a_L}{2\omega}\left[\bm{U}_L^{a-}\right]^\dagger\bm{\Gamma}_L^{00}\bm{U}_L^{a-}\\
\bm{V}_R^+=\frac{a_R}{2\omega}\left[\bm{U}_R^{r+}\right]^\dagger\bm{\Gamma}_R^{00}\bm{U}_R^{r+}
\end{align}
where $\bm{\Gamma}_L^{00}=i(\bm{\Sigma}_L^{00}-\bm{\Sigma}_L^{00\dagger})$ and $\bm{\Gamma}_R^{00}=i(\bm{\Sigma}_R^{00}-\bm{\Sigma}_R^{00\dagger})$. The velocity matrices in Eq. (\ref{eq:velocityMatrix}) are diagonal matrices with the diagonal matrix elements equal to the group velocities associated with the eigenvectors in $\bm{U}_L^{a-}$ and $\bm{U}_R^{r+}$. Finally, we have the expression for the $N_R\times N_L$ transmission matrix:
\begin{align}\label{eq:transmissionMatrix}
\bm{t}=\frac{2i\omega}{\sqrt{a_La_R}}\sqrt{\bm{V}_R^{r+}}\left[\bm{U}_R^{r+}\right]^{-1}\bm{G}_{RL}^r\left[\bm{U}_L^{a-\dagger}\right]^{-1}\sqrt{\bm{V}_L^{a-}}
\end{align}
where
\begin{align*}
\bm{G}_{RL}^r = \bm{g}_R \bm{H}_{RD} \bm{G}_D \bm{H}_{DL} \bm{g}_L        
\end{align*}
The square modulus of the matrix element $|t_{m,n}|^2$ represents the proportion of energy converted in transmission to the $m^\mathrm{th}$ wavemode in the right lead from the $n^\mathrm{th}$ wavemode in the left lead. We note here that the expression in Eq. (\ref{eq:transmissionMatrix}) efficiently yields  \emph{all} the possible transmission amplitudes in one computational step. The reflection matrix can be similarly computed (in the left lead):
\begin{align}\label{eq:reflectionMatrix}
\bm{r}=\frac{2i\omega}{a_L}\sqrt{\bm{V}_L^{r-}}\left[\bm{U}_L^{r-}\right]^{-1}\left(\bm{G}_{L}^r-\bm{Q}_L^{-1}\right)\left[\bm{U}_L^{a-\dagger}\right]^{-1}\sqrt{\bm{V}_L^{a-}}
\end{align}
In the above, $\bm{Q}_L^{-1}$ is the bulk Green's function in the left lead with:
\begin{align}
\bm{Q}_L=(\omega^2+i\eta)\bm{I}_L-\bm{H}_L^{00}-\bm{H}_L^{10}\bm{g}_{L-}^{r}\bm{H}_L^{01}-\bm{H}_L^{01}\bm{g}_{L+}^{r}\bm{H}_L^{10}
\end{align}
For a better interpretation in the results section, a set of new matrices $\tilde{\bm{t}}$, $\tilde{\bm{r}}$ can be defined as $\tilde{\bm{t}}=\bm{t}^\dagger\bm{t}$ and $\tilde{\bm{r}}=\bm{r}^\dagger\bm{r}$. Here, the diagonal values $\tilde{\bm{t}}$ and $\tilde{\bm{r}}$ ($\tilde{\bm{t}}_{nn}$, $\tilde{\bm{r}}_{nn}$) represent the total normalized transmitted and reflected energy respectively when the incident mode is mode $n$. Flux conservation is now given by the straightforward relation:

\begin{equation}
    \label{Encons}
    \tilde{\bm{t}}_{nn}+\tilde{\bm{r}}_{nn}=1;\quad \forall n
\end{equation}

\begin{figure}[htp]
\centering
\includegraphics[scale=0.6]{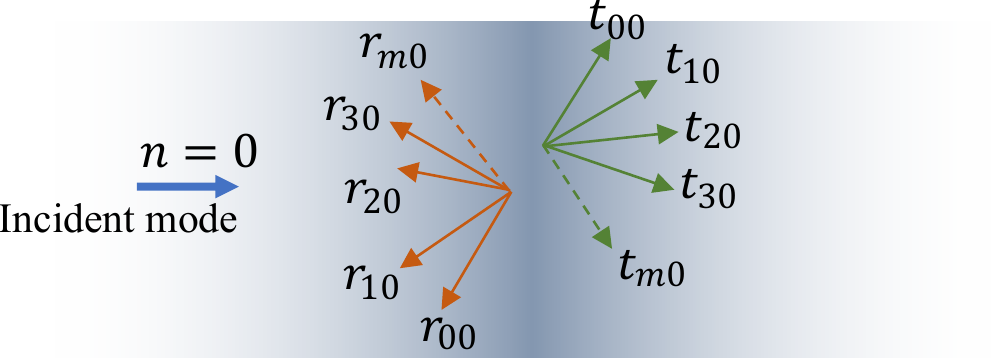}
\caption{Scattering of a single mode from a device/defect in a waveguide. The figure shows an incoming mode $0$ wave from left which excites reflected and transmitted modes before and after the heterogeneity. $r_{i0},t_{i0}$ are showing scattered modes caused by mode zero of the incident field.}
\label{energy}
\end{figure}

Figure (\ref{energy}), schematically shows how an incident mode breaks into fractions of reflected and transmitted modes.

\subsubsection{Scattering from defects in a waveguide}\label{sec:sccdefect}
\begin{figure}[htp]
\centering
\includegraphics[scale=0.60]{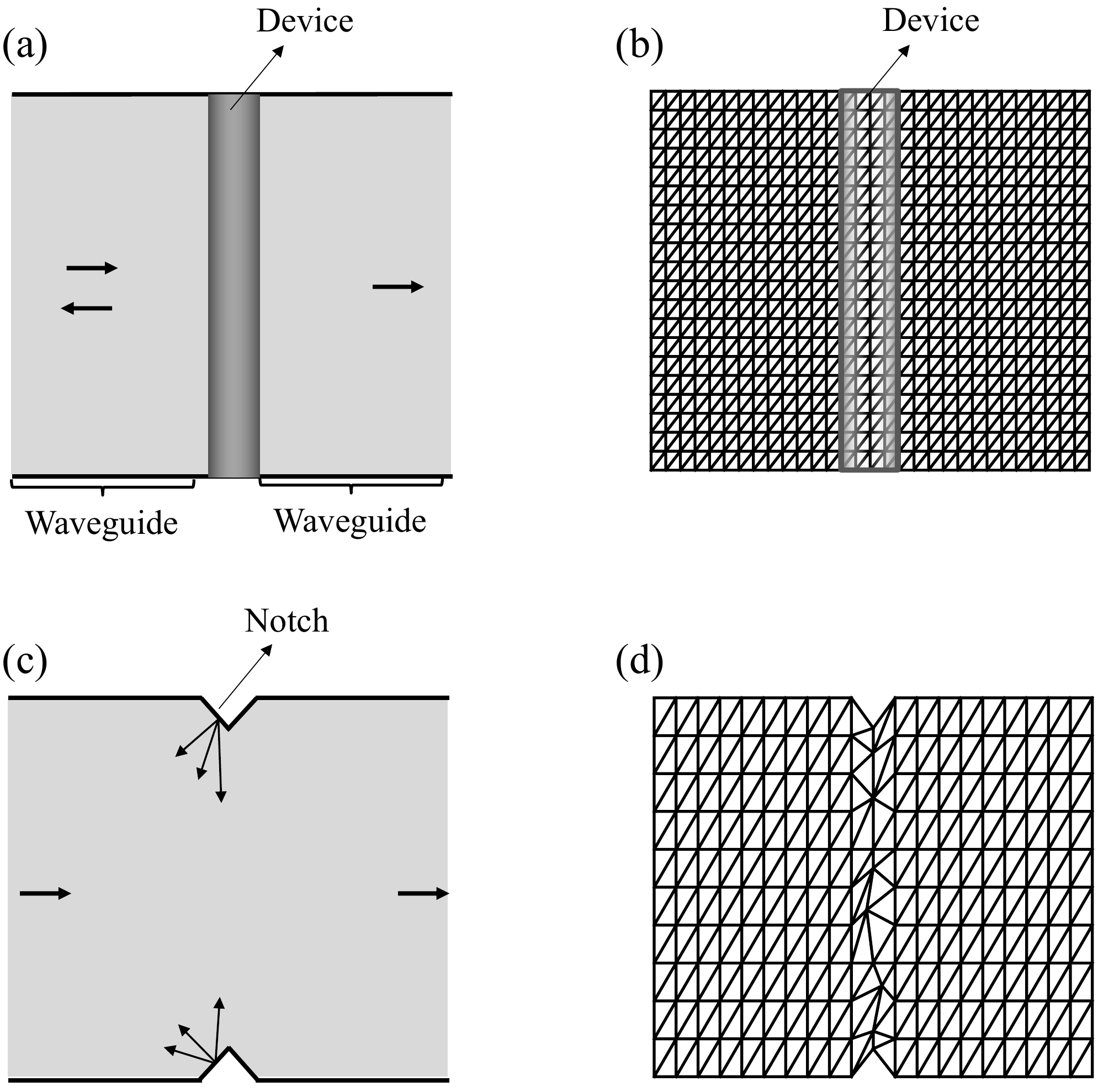}
\caption{Schematic of a 2D elastic waveguide with a defect. (a) and (b) show the general configuration and the mesh when the defect is in the form of a region of material heterogeneity. (c) and (d) show the same when the defect is in the form of a notch.}
\label{defectiveWG}
\end{figure}

We now consider two examples of defects in a waveguide. The two cases are shown in \fref{defectiveWG} which also shows the meshes. The first defect, shown in \fref{defectiveWG}a and b, is in the form of a region that differs from the rest of the waveguide only in terms of material properties. In this case, it is easy to create a uniform grid that naturally lends itself to AGF computations. In the second example, shown in Figs. \ref{defectiveWG}c and d,  we consider a notch-shaped defect in the waveguide. In this case, a uniform grid cannot be applied everywhere.

It is worth noting that the current version of the AGF used in this paper only requires translational invariance in the leads, while the mesh shape and arrangement in the scatterer can be arbitrary. Specifically, the analytical and numerical calculations of the surface Green's matrices are the only instances in which adjacent blocks (so the mesh) are required to be homogeneous in space. This feature of the AGF allows us to tackle more complicated scatterers with arbitrary meshes. A good example of this is the scattering problem in a waveguide with a notch, shown in \fref{defectiveWG}, where an arbitrarily meshed scatterer is connected to two rectangular meshes representing the environment.

\begin{figure}[htp]
\centering
\includegraphics[scale=0.27]{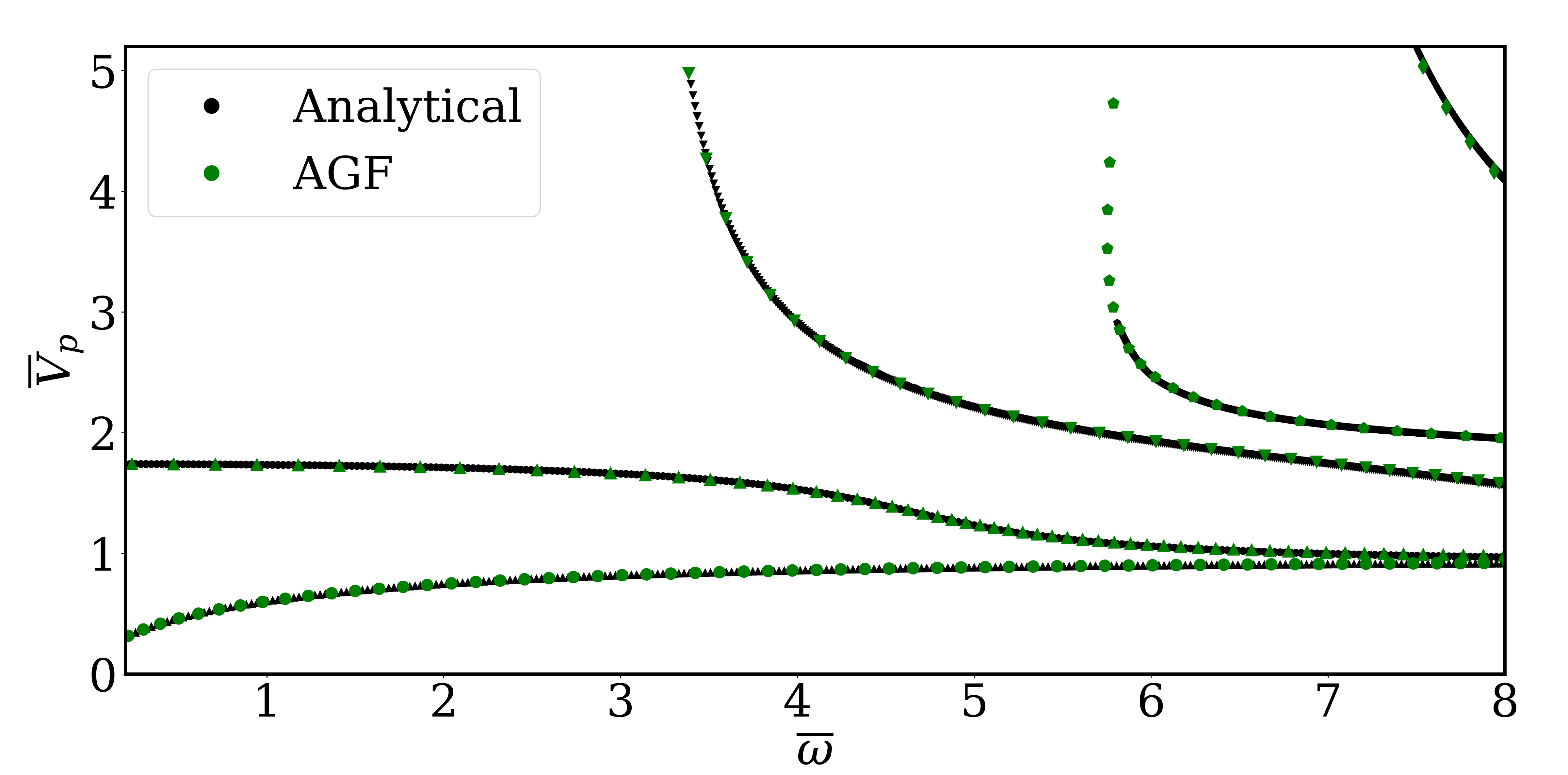}
\caption{The dispersion relation between normalized phase velocities and normalized angular frequencies for the five first modes in the bulk of the waveguide.}
\label{Dispersion}
\end{figure}

We now present two numerical examples to illustrate the formulation presented above for the scattering and dispersion problem involving waveguides. For the 2D waveguide problem, we use a FEM code written in Python using the FEniCS toolbox to obtain the discretized domain, stiffness, and mass matrices, which are then used in further calculations. The first example involves in-plane scattering in an elastic waveguide with a material discontinuity acting as a scatterer (\fref{defectiveWG}a, b). In the second example, everything is the same except that the scatterer is modeled as a notch. For the environment, we consider mechanical properties of $E= 8$ GPa, $\rho=1180$ kg/$m^3$, and $\nu=0.34$. For the case where the scatterer is a material discontinuity, we use $E= 300$ GPa, $\rho=8000$ kg/$m^3$, and $\nu=0.27$ as the mechanical properties of the scatterer. For this same case study, the length of the scatterer, the width of both the scatterer and waveguide, and the horizontal and vertical length scales of the rectangular mesh are also $2.46 \times 10^{-2}$ m, $8 \times 10^{-2}$ m, $3 \times 10^{-3}$ m, and $4 \times 10^{-3}$ m, respectively. The horizontal and vertical length scales in the rectangular mesh of the waveguide with a notch are $4 \times 10^{-3}$ m and $8 \times 10^{-3}$ m, respectively. \fref{Dispersion} shows the dispersion curves for the first five modes in the bulk of the waveguide, generated using equation (\ref{eq:bulkDispersion}). Here, $\overline{V}_p=c/V_s$ and $\overline{\omega}=D\omega/V_s$, where $c$ is the phase velocity, $V_s$ is the shear wave velocity of the medium, and $D$ is the width of the waveguides. These results have been compared with the solutions of the Lamb wave frequency spectrum \cite{lamb1917waves} and show good agreement.

\begin{figure}[htp]
\centering
\includegraphics[scale=0.18]{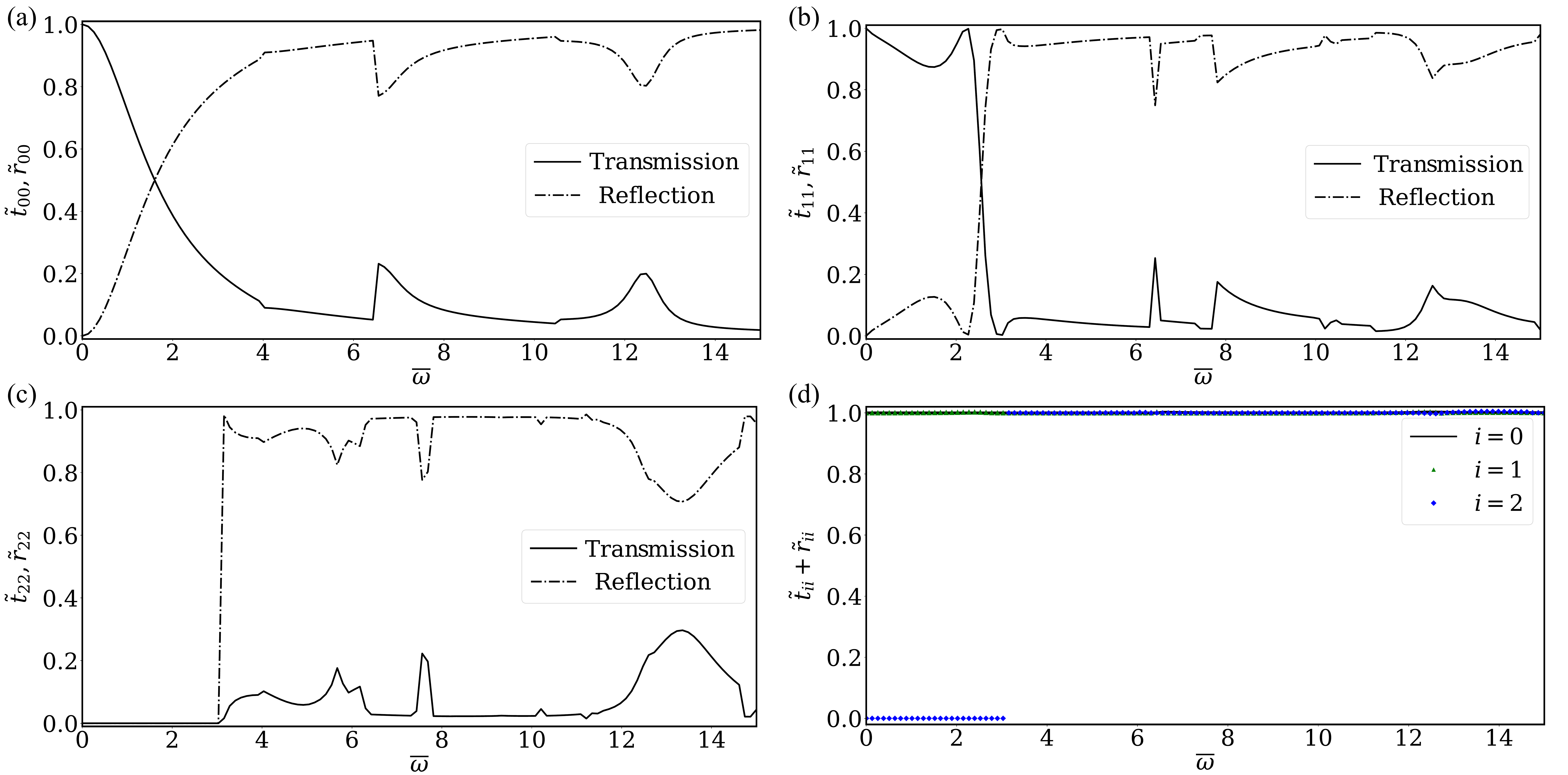}
\caption{Results for material heterogeneity defect in a waveguide. Plots (a), (b), and (c) depict the sum of the squared transmission/reflection coefficients for incident modes 0, 1, and 2 respectively. Plot (d) shows the flux conservation.}
\label{Fig:sccen}
\end{figure}

In \fref{Fig:sccen}, some scattering results along with the satisfaction of energy conservation are presented in the frequency domain. \fref{Fig:sccen}a shows the sum of the squared transmitted/reflected coefficients for mode zero of the incident wave. \fref{Fig:sccen}b and c show the corresponding values for modes 1 and 2 of the incident wave, respectively. The inherent law of energy conservation is also checked for the first three modes of incidence in \fref{Fig:sccen}d. The zero transmission/reflection induced by the third mode incident wave up to $\overline{\omega}\approx 3.15$ ($\omega \approx 6\times 10^4$) shown in \fref{Fig:sccen}c occurs because there is no third propagating mode within the mentioned frequency range, as can be confirmed by the dispersion curves in \fref{Dispersion}. The figure evaluating energy conservation (\fref{Fig:sccen}d) also shows zero reflected and transmitted energy fractions.

\begin{figure}[htp]
\centering
\includegraphics[scale=0.18]{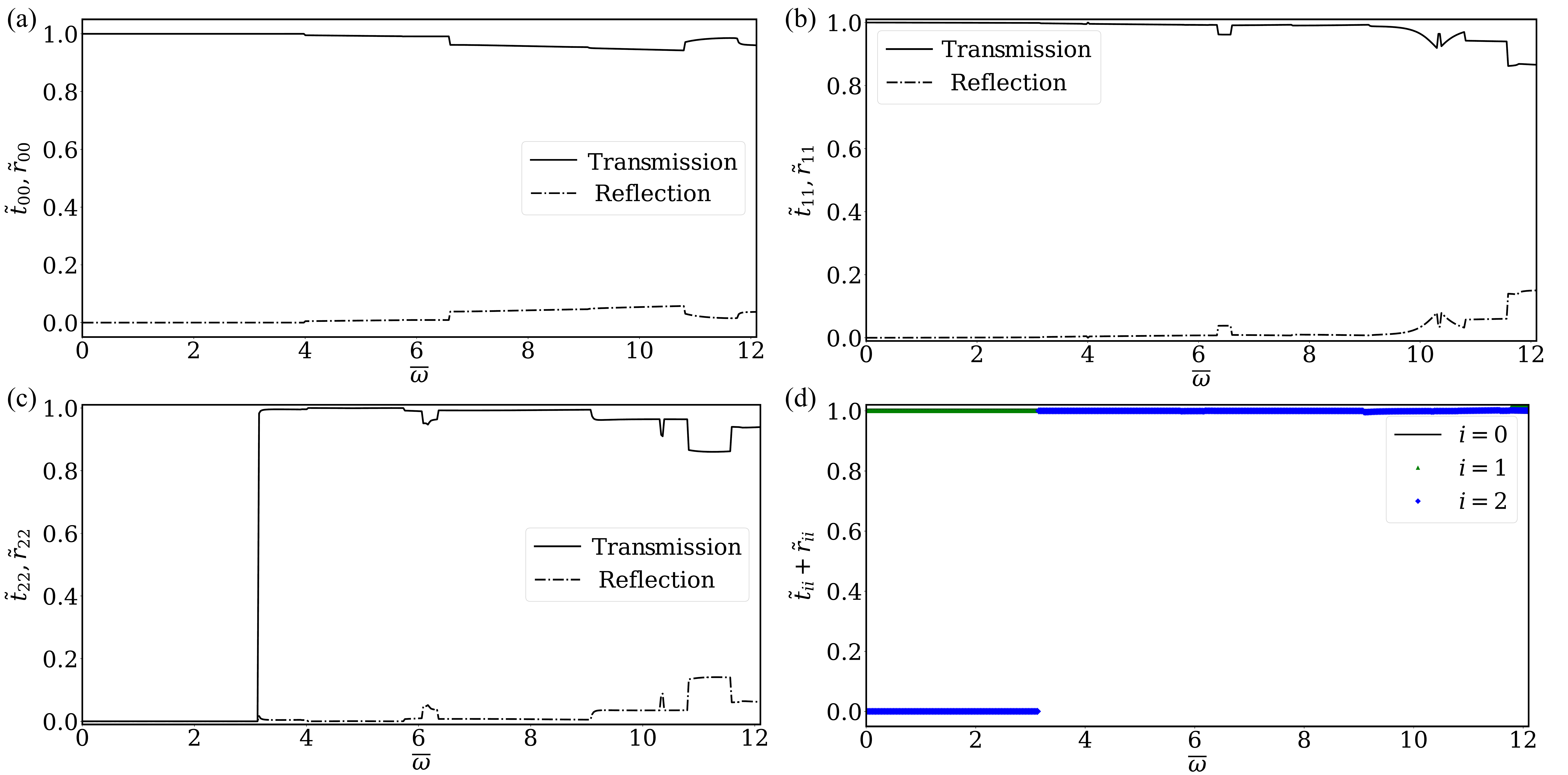}
\caption{Results for a notch defect in a waveguide. Plots (a), (b), and (c) depict the sum of the squared transmission/reflection coefficients for incident modes 0, 1, and 2 respectively. Plot (d) shows the flux conservation.}
\label{Fig:sccenN}
\end{figure}

Figure (\ref{Fig:sccenN}) shows the transmission/reflection calculation results for a waveguide with a notch, presented for individual incident modes 0, 1, and 2. As before, the conservation of the energy flux entering and exiting the scattering zone is checked in \fref{Fig:sccenN}d. Again, the initial zero transmission/reflection spectrum, happening for modes $2$ and beyond, are obeying the non-propagating nature of those modes in the low-frequency regime which can be verified by the dispersion curve of the homogeneous medium, \fref{Dispersion}. Following the capability of the AGF method for accepting any arbitrary mesh in the scatterer zone (explained in section \ref{sec:sccdefect}), the scattering results in \fref{Fig:sccenN} were run for different mesh distributions of the device, and the results all showed an excellent match.

\section{Conclusions}
Following the perspective of ``open systems'' which proposes the viewpoint of ``focusing on the scatterer and the excitations from the environment on the contact DOFs'' to solve scattering problems, in this paper we applied the atomistic green's function (AGF) method to find the effective Hamiltonians in one and two-dimensional problems more systematically. Through the use of the AGF method, we demonstrated how it is possible to reduce an infinite dimensional scattering problem to a finite problem. This allows us to derive a finite-dimensional effective Hamiltonian and Green's matrix for the scatterer, which not only captures the properties of an isolated scatterer but also contains information about its interaction with the surrounding environment. We also showed that the effective Hamiltonian derived in this manner allows the scattering solution to be independent of any far-field non-reflecting boundary conditions. Having only local interactions, we were able to calculate finite-dimensional surface Green's matrices analytically and numerically for 1D and 2D problems, avoiding the need to deal with infinite-dimensional Green's matrices for each lead. The immediate benefit of having surface Green's matrices was the straightforward calculation of the self-energies, which represent the interaction between the device and its environment. We also used the Caroli formulae for 1D problems and mode matching for 2D problems to calculate the scattering coefficients for both 1D and 2D scattering scenarios.
In this paper, we presented numerical results for two different scattering scenarios: a 1D case with device heterogeneity and a finite phononic crystal, and a 2D case with device heterogeneity and a notch-shaped defect. These examples showcase the versatility and effectiveness of the AGF method in tackling a variety of scattering problems.\par
The scattering solution obtained through the AGF method not only allows us to bypass the concern of far-field boundary conditions, but also provides valuable insight for future research on inverse problems involving the design of scatterers. By examining the relationship between the scattering solution and the effective Hamiltonian, we can gain a deeper understanding of the underlying physical processes at play and apply this knowledge to devise more effective scatterers.

\section{Appendix}
\subsection{Decimation Technique}\label{appdec}

\begin{figure}[htp]
\centering
\includegraphics[scale=0.6]{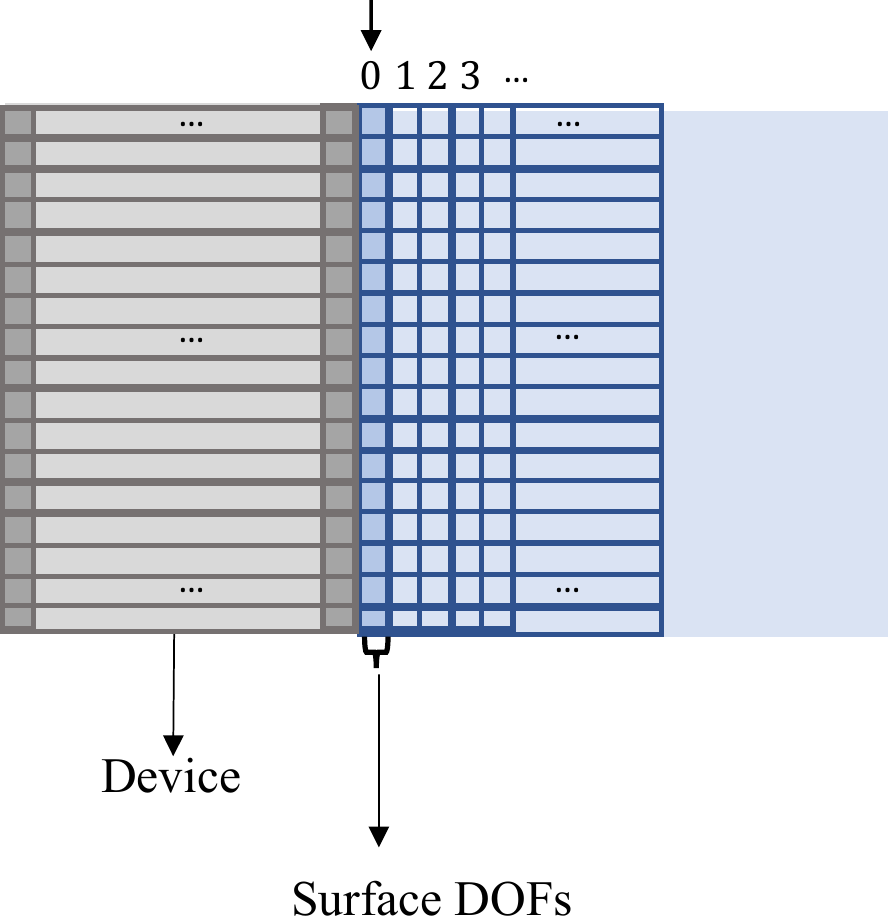}
\caption{A schematic of the FEM model. The arrow on top is showing those degrees of freedom that are effective in Green's function of the right lead.}
\label{Fig:Decimation}
\end{figure}

In this appendix, we provide a brief introduction to a numerical technique called Decimation, which can be used to calculate the surface Green's matrices of the left and right leads. The Decimation technique is based on the general equation for the Green's matrix, which is given by $\left[\left(\omega^2+i\eta\right)\bm{I}-\bm{H}_\alpha\right]\bm{g}_\alpha=\bm{I}$, where $\alpha=R$ or $L$. The technique takes advantage of the fact that only one block of the Green's matrix is of interest, depending on the chosen lead. In this appendix, we focus on the right lead and use the convention introduced in Eq. (\ref{eq:HBandedExp}) to expand the equation for the Green's matrix of the right lead as:

\begin{align}\label{gmR}
\nonumber 
&m=0;\quad \left[\left(\omega^2+i\eta\right)\bm{I}-\bm{H}_{R}^{00,s}\right]\bm{g}_{R}^{00}-\bm{H}_{R}^{01}\bm{g}_{R}^{10}=\bm{I}\\
\nonumber
&m=1;\quad-\bm{H}_{R}^{10}\bm{g}_{R}^{00}+\left[\left(\omega^2+i\eta\right)\bm{I}-\bm{H}_{R}^{00}\right]\bm{g}_{R}^{10}-\bm{H}_{R}^{01}\bm{g}_{R}^{20}=\bm{0}\\
\nonumber 
&\vdots\tag{6.1.1}\\
\nonumber 
&m=m;\quad-\bm{H}_{R}^{10}\bm{g}_{R}^{(m-1)0}+\left[\left(\omega^2+i\eta\right)\bm{I}-\bm{H}_{R}^{00}\right]\bm{g}_{R}^{m0}-\bm{H}_{R}^{01}\bm{g}_{R}^{(m+1)0}=\bm{0}
\end{align}
where $\bm{H}_R^{10}$ and $\bm{H}_R^{01}$ are the coupling matrices between degrees of freedom in two successive columns (as shown in Fig. (\ref{Fig:Decimation})), and the set of equations is simply the result of matrix multiplication of all the rows of $\bm{H}_R$ in the first column of $\bm{g}_R$. It is known that the $\bm{g}_R^{00}$ block of the Green's matrix of the right lead represents the \textit{response at the surface degrees of freedom to the excitation on surface degrees of freedom,} as shown by the downward arrow in Fig. (\ref{Fig:Decimation}). More generally, $\bm{g}_R^{mn}$ represents \textit{the response at the degrees of freedom on the $m^{th}$ column to the excitation on the degrees of freedom on the $n^{th}$ column.} The Decimation technique suggests substituting $\bm{g}_R^{i0}$, where $i=2k+1$ and $k\in \mathbb{Z}_{\geq 0}$, with expressions found in terms of $\bm{g}_R^{j0}$, where $i=2k$ and $k\in \mathbb{Z}_{\geq 0}$, for the first step. The updated form of Eq. (\ref{gmR}) for $m=0$ becomes:

\begin{equation}
    \label{form0}
\nonumber    
\left\{\left(\omega^2+i\eta\right)\bm{I}-\bm{H}_{R}^{00,s}-\bm{H}_{R}^{01}\left[\left(\omega^2+i\eta\right)\bm{I}-\bm{H}_{R}^{00}\right]^{-1}\bm{H}_{R}^{10}\right\}\bm{g}_{R}^{00}-\bm{H}_{R}^{01}\left[\left(\omega^2+i\eta\right)\bm{I}-\bm{H}_{R}^{00}\right]^{-1}\bm{H}_{R}^{10}\bm{g}_{R}^{20}=\bm{I}\tag{6.1.2}
\end{equation}

In Eq. (\ref{gmR}), the line for $m=0$ was describing a relation between the surface Green's matrix ($\bm{g}_R^{00}$) and $\bm{g}_R^{10}$. The best interpretation says: given an impulse excitation at the left-most DOFs, the first line of Eq. (\ref{gmR}) was describing the relation between the response at the DOFs on $0^{th}$ column and the response at the DOFs of the  $1^{st}$ column. However, the updated version, Eq. (\ref{form0}), is showing a similar concept, but between the $0^{th}$ column and a further column. The locality of the interactions in this FEM model mandates that the new coupling term, $\bm{H}_{R}^{01}\left[\left(\omega^2+i\eta\right)\bm{I}-\bm{H}_{R}^{00}\right]^{-1}\bm{H}_{R}^{10}$, should be smaller compared to its initial counterpart. Doing the same thing for other lines of equations in Eq. (\ref{gmR}) gives similar updated versions, in which the relations are between further DOFs with smaller coupling terms. The decimation technique uses the same logic and updates the equations in Eq. (\ref{gmR}) iteratively, in a way that at each step, it is giving relations between the DOFs of two further columns with smaller coupling terms. After enough iterations, the equation for the surface green's matrix of the right lead can be written as:

\begin{equation}
    \label{surfR}
    \nonumber
\left[\left(\omega^2+i\eta\right)\bm{I}-\bm{H}_{R}^{00,s'}\right]\bm{g}_{R}^{00}-\bm{\tau}\bm{g}_{R}^{P0}=\bm{I}\tag{6.1.3}
\end{equation}

where $P$ is pointing to a far enough column, $\bm{H}_{R}^{00,s'}$ is representing all the updates on $\bm{H}_{R}^{00,s}$ and, $|\bm{\tau}|\rightarrow 0$. Then the surface green's matrix can be approximated as:

\begin{equation}
    \label{surfRf}
    \nonumber
    \bm{g}_{R}^{00}=\left\{\left[\left(\omega^2+i\eta\right)\bm{I}-\bm{H}_{R}^{00,s'}\right]\right\}^{-1}\tag{6.1.4}
\end{equation}

\section*{Acknowledgments}
AS acknowledges support from the US National Science Foundation award \#2219203. ZYO acknowledges support for this work by A*STAR, Singapore with funding from the Polymer Matrix Composites Program (SERC Grant No. A19C9a0044).

\section*{References}

%

\end{document}